\documentclass[reprint,superscriptaddress,aps,pra]{revtex4-2}

\usepackage{amssymb}
\usepackage{amsmath}
\usepackage{ulem}
\usepackage{graphicx}

\usepackage[utf8]{inputenc}
\usepackage[english]{babel}
\usepackage[T1]{fontenc}
\usepackage{textcomp}
\usepackage{graphicx}

\usepackage[dvipsnames]{xcolor}

\usepackage[colorlinks,allcolors=blue]{hyperref}

\makeatletter
\def\@hangfrom@section#1#2#3{\@hangfrom{#1#2}#3}
\def\@hangfroms@section#1#2{#1#2}
\makeatother

\begin{document}

\author{A.\ A.\ Fomin} 
\author{M.\ Yu.\ Petrov}
\affiliation{Spin Optics Laboratory, St.~Petersburg State University, 198504 St.~Petersburg, Russia}

\author{A.\ S.\ Pazgalev}
\affiliation{Ioffe Institute, Russian Academy of Sciences, St.~Petersburg, 194021, Russia}

\author{G.\ G.\ Kozlov}
\author{V.\ S.\ Zapasskii}
\affiliation{Spin Optics Laboratory, St.~Petersburg State University, 198504 St.~Petersburg, Russia}

\title {Optically driven spin-alignment precession}

\begin{abstract} 
The effect of optically driven spin precession discovered by Bell and Bloom [W.~E.\ Bell and A.~L.\ Bloom, Phys.\ Rev.\ Lett.\ \textbf{6}, 280 (1961)] 
is widely used nowadays as a basis for numerous experiments in fundamental physics and for diverse applications. In this paper we consider a much less popular version of the light-induced spin precession that does not imply coherent precession of the spin-system magnetization and is excited by linearly (rather than circularly) polarized light. Pump-probe measurements performed on the $D_2$ line of cesium vapor show that the magnitude of the signal of the optically driven spin-alignment precession, in ``vacuum'' cells (with no buffer gas) is close to that of classical spin-orientation precession. In the presence of buffer gas, however, the signal of spin-alignment precession appears to be strongly suppressed.  The discovered effect is ascribed to spin mixing of excited states of cesium atoms in the cycle of optical pumping. 
\end{abstract}

\maketitle

\section{Introduction}

Optical orientation, developed by Kastler \cite{Kastler}, was widely used for more than 40 years in fundamental and applied research of atomic and solid-state paramagnets~\cite{OO}.
The method is based on the transfer of angular momentum from circularly polarized light to the spin system. From the viewpoint of symmetry, the effect of optical orientation has much in common with the inverse Faraday effect and other effects of optically induced magnetization~\cite{Ziel,Pershan,Ziel2,Rand}. The process of optical orientation performed, as a rule, in a magnetic field aligned along the light beam propagation (the Faraday configuration) affects magnetization of the material and its optical activity.\looseness=-1

A more interesting effect of optical orientation can be observed in a transverse magnetic field (Voigt geometry) that causes the light-induced angular momentum to precess around the field direction \cite{BB1}. Under these conditions, optical orientation can be achieved using the light beam modulated at the precession frequency (Larmor frequency). This effect is often referred to as optical orientation in the rotating frame or optically detected magnetic resonance \cite{BB2,beam}. In the latter case, the modulated light affects the spin system as an effective \textrm{ac}\ magnetic field driving the spin precession in  conventional electron paramagnetic resonance spectroscopy. Due to the high finesse of the spin precession resonance in atomic systems (classical objects of this technique) in combination with the compactness of the apparatus and the possibility of operating at room temperature and in low magnetic fields, the effects of optically driven spin precession have gained wide popularity as a tool of research and magnetometric applications \cite{budker}. 

In the above effects, the symmetry of the light-induced anisotropy brought by the circularly polarized light (the symmetry of \textrm{axial vector} or \textrm{pseudovector}) explains the fundamental possibility of its conversion to magnetization.  There is another effect of the light-induced arrangement of paramagnetic particles (spins) that does not imply any magnetization of the medium. This is the effect of so-called \textrm{spin alignment} induced by linearly polarized light and revealed as linear (rather than circular) anisotropy of the paramagnet~\cite{Happer1}. When the distinguished direction of the optical perturbation (for the optically induced alignment, this is the direction of the electric-field vector of the light wave) does not coincide with that of the magnetic field applied to the sample, the light excites a nonstationary (superposition) state of the system (exactly like circularly polarized light does in the Voigt geometry). As a result, the axis of alignment of the spin system in the magnetic field, in the general case, should precess around the field, giving rise to oscillating linear anisotropy (birefringence and dichroism) of the medium.  In full analogy with the optically driven spin-orientation precession described above, the modulated linearly polarized light can excite spin-alignment precession under resonant conditions.  This version of the optically driven spin precession differs in many respects from the above optically detected magnetic resonance  in circularly polarized light. In particular, in the Faraday configuration, it is observed at the double Larmor frequency and is not accompanied by precessing magnetization of the spin system. For the latter reason, this type of precession evidently cannot be excited with the aid of a resonant \textrm{magnetic field}  at the double Larmor frequency.  Specific features of the spin-alignment precession do not preclude its use in fundamental research, but altogether this experimental approach is rarely employed nowadays for applied purposes \cite{Bevil, Rom, Gil, Ros}.\looseness=-2

It is noteworthy that the problem of the spin-precession resonance width was always of interest not only from the viewpoint of physics of the effect but also (and even primarily) from the viewpoint of its magnetometric applications, when the finesse of the resonance directly controls the sensitivity of the device. It was found that the effect of collisions of atoms with the cell walls, highly destructive for the spin polarization, can be strongly suppressed by adding a gas of rare atoms (buffer gas) \cite{Brossel, Dehmelt,Fran} or by using the cells with paraffin-like coatings \cite{Bou1, Bou2, Bou3}. In several applications of the spin-alignment resonance, only the second method was used \cite{Al1, Al2, Al3, Al4}. 

In this paper we present an experimental study of spin-alignment precession induced in cesium vapor by resonant linearly polarized light. We consider specific polarization properties of the effect, describe optimal conditions for its observation at the fundamental harmonic of the Larmor frequency, and compare the effects of buffer gas on the optically induced spin-alignment and spin-orientation precession. The discovered suppression of the optically driven spin-alignment precession in the presence of a buffer gas is ascribed to the effect of spin mixing in the excited optical state and the absence of depopulation pumping.\looseness=-1

The paper is organized as follows. In Sec.~\ref{Sec_General} we present a general description of experimental schemes suitable  for excitation and detection of spin-orientation and spin-alignment precession at the Larmor frequency. Section~\ref{Sec_Experimental} depicts the experimental setup and details of  the measurements. Experimental results, demonstrating specific features of the optically induced spin-alignment precession effect, compared with those of the spin-orientation precession, are presented in Sec.~\ref{Sec_Results}. In Sec.~\ref{Sec_Discussion} we discuss the discovered distinctions between the effects of spin-alignment and spin-orientation precession in buffer-gas-filled cells. The paper ends with a few concluding remarks.\looseness=-1

\section{General considerations}
\label{Sec_General}

A common characteristic feature of spins in external magnetic field is their precession. Since, in the equilibrium spin system, the phases of this precession for different atoms are random, this spontaneous precession is not revealed in any regular form and can be observed only as a peak in the magnetization noise (or Faraday rotation noise) spectrum \cite{Zap}. The situation changes when the spin precession of individual spins is synchronized by some external perturbation, and the medium starts to exhibit oscillating behavior. 

\begin{figure}
\includegraphics[width=.95\columnwidth,clip]{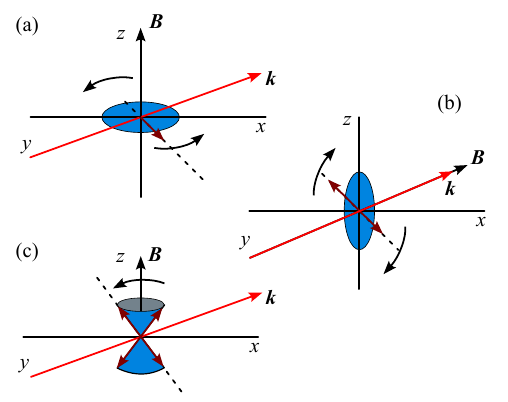}
\caption{Three geometries for detecting optically driven spin precession: (a) Voigt geometry to observe the spin-orientation precession at the Larmor frequency, (b) Faraday geometry to observe the spin-alignment precession at the double Larmor frequency, and (c) Voigt geometry to observe the spin-alignment precession at the Larmor frequency. Here $k$ is the light wave vector and $B$ is the magnetic field. Axes of the precessing anisotropy are indicated by dashed lines. The surfaces swept by the precessing spins are blue. %
}\label{fig1} 
\end{figure}

The simplest coherent spin precession, which can be excited by the modulated circularly polarized light or by an oscillating magnetic field, implies precession of spin orientation accompanied by oscillating magnetization of the system at the Larmor frequency [Fig.\ \ref{fig1}(a)].  From the viewpoint of optics, such a precessing axial vector corresponds to the precessing gyrotropy of the medium (the precessing antisymmetric part of its optical susceptibility). This means that, in the Voigt geometry [Fig.\ \ref{fig1}(a)], which is considered to be most convenient for observation of the spin-orientation precession, the Faraday rotation of the probe beam oscillates at the Larmor frequency. 

It is important that the symmetry of the spin alignment, which is characterized by a chosen axis of spin orientation, with no definite sign of chirality, is essentially different. The most convenient geometry for detecting the spin-alignment precession is that of Faraday, shown in Fig.\ \ref{fig1}(b). The precessing spin alignment that can be induced by modulated linearly polarized light is revealed as a rotating symmetric part of the optical susceptibility tensor or as a rotating linear anisotropy. As is evident from symmetry properties of the alignment (and of the linear anisotropy), rotation of its axis by $180^\circ$ returns it to its initial state. This means that the precessing alignment, in this geometry, can be detected only at the double Larmor frequency.
  
One can see, however, that when the axis of the alignment makes some angle $\theta$ with the magnetic-field direction, other than $90^\circ$ and $0^\circ$, the positions of the alignment axes of the system, in a half-period of the Larmor precession, for the light beam in the Voigt geometry, cease to be equivalent. Specifically, for the spin alignment excited at $\theta  = 45^\circ$, its axis, while precessing around the magnetic field, will change its direction by $90^\circ$ each half-period of the precession [see Fig.\ \ref{fig1}(c)]. As a result, the relevant linear anisotropy of the medium, under these conditions, will oscillate at the fundamental Larmor frequency. These regulations of the spin-alignment precession in a different experimental arrangement are also revealed in the spin-alignment noise experiments~\cite{Zap1}. 

Thus, we conclude that, from the viewpoint of symmetry considerations, the effects of spin-orientation precession and spin-alignment precession can be observed in the Voigt geometry at the fundamental harmonic of the Larmor frequency.      

\section{Experimental}
\label{Sec_Experimental}

\begin{figure}
\includegraphics[width=.95\columnwidth,clip]{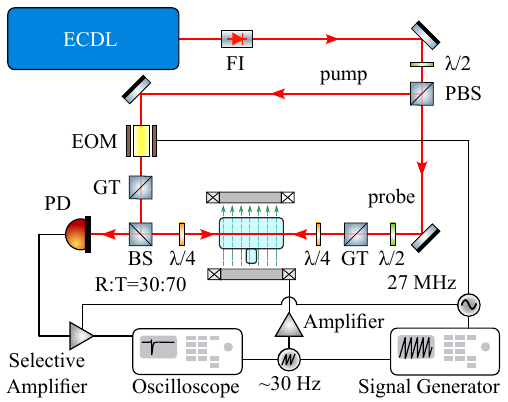}
\caption{Schematic of the experimental setup. See the text for acronym definitions.%
}\label{fig2} 
\end{figure}

A schematic of the experimental setup is shown in Fig.\ \ref{fig2}. As a light source, we used the external-cavity diode laser  (ECDL) operated in a free-run regime and tunable within the range of the $D_2$ line of cesium. The output emission of the laser, after passing through the Faraday isolator (FI), was split by the $\lambda/2$ waveplate and polarizing beam splitter (PBS) into pump and probe beams of approximately equal intensity. The pump beam, intensity modulated at a frequency of $27$~MHz, after passing through the non-polarizing beam splitter (BS), was directed to the cell with cesium vapor.  The light intensity modulator was comprised of an electro-optical modulator (EOM) followed by the Glan-Taylor polarizer (GT). The amplitude of the voltage applied to the EOM was relatively small (much smaller than the half-wave voltage) and the modulation depth of the probe beam intensity was correspondingly small (around 10\%). The linearly polarized probe beam, attenuated by the $\lambda/2$ waveplate and GT prism, passed through the cesium cell in the opposite direction and, after passing through the BS, hit the photodetector (PD). This geometry of counter propagating beams did not pursue the goal of canceling the Doppler broadening. It was chosen to extinguish the light intensity modulation  in the probe channel with the maximum  overlap of the two beams over the cell. In this study we ignored  possible effects of the Doppler-free geometry on the spectral behavior of the optically driven precession.

The cell with cesium heated to approximately $60^\circ$C was placed to the magnetic field directed across the light beam propagation and modulated in magnitude (by sawtooth law) at a frequency of approximately $30$~Hz. The output signal of the PD was selectively amplified at the frequency of the pump-intensity modulation and fed to the input of the oscilloscope whose sweep was synchronized with the magnetic-field modulation. Thus, we could directly observe, on the screen of the oscilloscope, the intensity modulation signal as a function of the applied magnetic field. The signal arose at the point of the sweep where the light modulation frequency coincided with that of Larmor precession. The measurements were performed when both pump and probe beams passing through the cesium cell were polarized either linearly (with the same azimuth of the polarization plane) or circularly. Correspondingly, the detected signals in these two cases were related to the optically driven spin-alignment or spin-orientation precession. In the latter case, the quarter waveplates ($\lambda/4$) were placed on both sides of the cesium cell. We used three cylindrical cells ($20$~mm inner diameter and $20$~mm long) with cesium and different content of buffer gas (neon) having a buffer-gas pressure of $0$, $2$, and $10$ Torr. 

The wavelength of the laser ($2$~mW in power and approximately $2$~mm beam size) was tuned approximately to the center of the long-wavelength component of the $D_2$ line (transition from the ground-state sublevel $F = 4$), where the magnitude of the signal was the greatest (the hyperfine structure of the excited state was not resolved).

\section{Results of the measurements}
\label{Sec_Results}

\begin{figure}
\includegraphics[width=.95\columnwidth,clip]{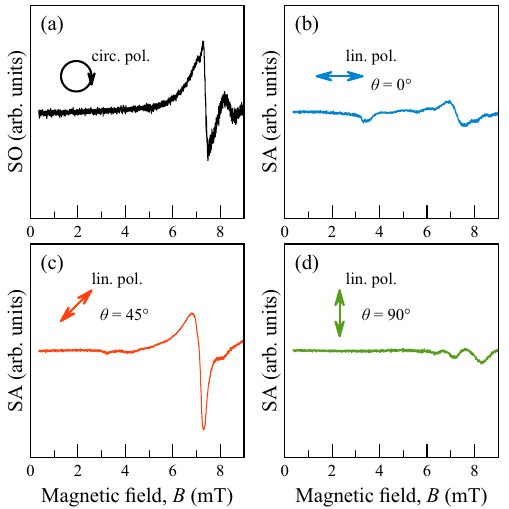}
\caption{(a) Spin-orientation (SO) precession resonance excited by circularly polarized light and (b)-(d) spin-alignment (SA) precession resonances excited by linearly polarized light with different azimuths of the polarization plane: (b) $\theta = 0^\circ$, (c) $\theta = 45^\circ$, and (d) $\theta = 90^\circ$.%
}\label{fig3} 
\end{figure}

The first experiments performed on the vacuum cell (with no buffer gas) were aimed at confirmation of the symmetry-based considerations, presented above. Polarizations of the pump and probe beams in these experiments were identical and either circular or linear with the same azimuths of the polarization plane. Figure\ \ref{fig3} shows the traces of the signal detected in circularly polarized light [Fig.\ \ref{fig3}(a)] and in linearly polarized light for three azimuths of the polarization plane: $\theta = 0^\circ$ [Fig.\ \ref{fig3}(b)], $\theta = 45^\circ$ [Fig.\ \ref{fig3}(c)], and  $\theta = 90^\circ$ [Fig.\ \ref{fig3}(d)]. 

We have found that, under these conditions, the spin precession resonances could be easily observed without any signal accumulation and that the signal of spin-alignment precession, as expected, was greatest in the vicinity of $\theta = 45^\circ$ and nearly vanished at $\theta = 0^\circ$ and $90^\circ$. For this reason, we performed all the measurements with optically driven spin-alignment precession at $\theta = 45^\circ$. One more result of these measurements, which we consider important, is that the magnitudes of the spin-alignment and spin-orientation signals, under these conditions, were nearly equal. This means that when choosing between these two effects for practical applications (e.g., in magnetometry), one can be guided more by their characteristic properties than by their sensitivities. 

\begin{figure}
\includegraphics[width=.95\columnwidth,clip]{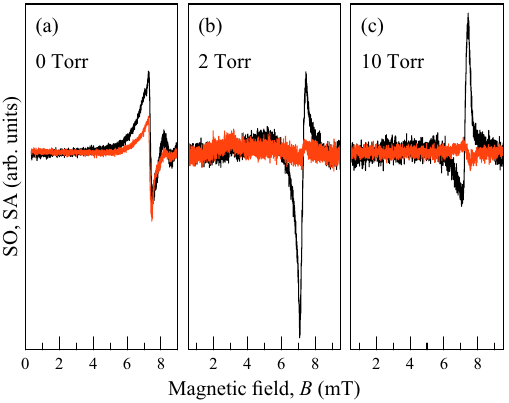}
\caption{Comparison of the spin precession signals  in Cs cells with different amount of buffer gas: (a) $0$\ Torr, (b) $2$\ Torr, and (c) $10$\ Torr. The black and red (light gray) curves correspond to spin-orientation (SO) and spin-alignment (SA) precession resonances, respectively.%
}\label{fig4} 
\end{figure}

The measurements performed on the cells with buffer gas show that the effect of buffer gas on the resonant precession signal is essentially different for the spin-alignment and spin-orientation precession and the equivalence of these two methods is violated. Figure~\ref{fig4} shows traces of the two signals in the cells with no buffer gas and with $2$ and $10$\ Torr of neon. As before, the signals of optically driven spin-alignment and spin-orientation precession were obtained in the Voigt geometry with linearly and circularly polarized light, respectively. One can see that, in the presence of buffer gas, the optical resonance of the spin-alignment precession practically vanishes, while the spin-orientation precession signal changes its sign (due to depolarization of the excited state \cite{Happer2}), but remains the same in magnitude. This destructive  effect of the buffer gas upon the spin alignment of cesium atoms, compared with its negligible effect on the spin orientation, deserves special consideration.\looseness=-1 

\section{Discussion}
\label{Sec_Discussion}
The effect of buffer gas on spin relaxation of alkali-metal atoms attracted considerable attention at the very earlier stages of development of the optical orientation effect~\cite{Fran}. This issue became especially important for optical magnetometry when the sensitivity of the device was directly governed by the spin resonance width. It was found that an oriented cesium atom may collide with a buffer-gas atom several hundred thousand times before it loses its ground-state  spin orientation. This fact was used both in optical magnetometry and in other applications of the spin-precession resonance to prolong the spin relaxation time in the buffer-gas-filled cells. At the same time, the effect of buffer gas on ground-state spin alignment was not so popular in spin physics and its applications were not given much attention (we did not manage to find any systematic research on the point). In most papers, the spin alignment was observed either in paraffin-coated cells or in cells with a small amount of buffer gas providing spectral resolution of the hyperfine structure of optical transitions.

Resonance excitation of magnetic resonance by means of light modulation is closely related to so-called coherent population trapping (CPT) and has much to do with the effects of frequency- and amplitude-modulated nonlinear magneto-optical rotation~\cite{Al1, GawlikAPL06}. The Bell-Bloom type of excitation produces a coherent optical waves, which excite long-lived ground state coherence of any kind. Most studies related to the frequency standards and magnetometers use orientation signals induced by circularly polarized light. It is known that for high gas pressures, where the hyperfine structure of the atomic absorption line is not resolved, the commonly used frequency modulated light cannot excite the CPT resonances \cite{Happer3}.

A rigorous treatment of the effect of buffer gas on the light-induced spin orientation and spin alignment requires consideration of a multitude of factors lying outside the scope of this paper. The observed effect is however of qualitative nature and we want to find its main cause and interpret it in a qualitative way. 

It is convenient to consider the resonance transverse pumping in a rotating frame. This is equivalent to longitudinal pumping by non-modulated light with no magnetic field. It will be shown that the probabilities of excitation by linearly polarized light are the same for all Zeeman ground-state levels if the upper optical state is not resolved. Since no population difference is produced, there is no reason to suppose that the alignment will arise with the modulated light by transverse pumping. The isotropic distribution will remain the same in either in a rotating or a laboratory frame.

We will show that, in small magnetic fields, no longitudinal alignment is created by the linearly polarized light with any orientation of its polarization vector with respect  to the magnetic field. Let us choose the quantization axis $z$ in the direction of the magnetic field, with the light propagating along the $y$~axis [Fig.\ \ref{fig1}(c)]. We assume that the electric-field vector of the linearly polarized light, ${\bf E}=\begin{pmatrix}E_x, & E_y, & E_z\end{pmatrix} = \mathop{\rm Re} \left[\begin{pmatrix}\sin\theta, & 0, &\cos\theta\end{pmatrix} e^{-i\omega t}\right]$, is aligned at an angle  $\theta$ to the magnetic field ($\omega$ is the optical frequency). For a circularly polarized field, $\mathbf{E}=\begin{pmatrix}-\sin\omega t, & 0, & \cos\omega t\end{pmatrix} = \mathop{\rm Re}\left[\begin{pmatrix} -i,& 0, & 1\end{pmatrix}e^{-i\omega t}\right]$. Hereafter, the light intensity is supposed to be  unity.

Decompose the electric field $\mathbf{E}_{LP} = \begin{pmatrix} E_+, & E_0, & E_-\end{pmatrix}$ in the basis of the circular parts 
$\mathbf{e}_+=-(i \mathbf{e}_x + \mathbf{e}_z)/\sqrt{2}$, $\mathbf{e}_0=\mathbf{e}_z$, and $\mathbf{e}_- = (-i \mathbf{e}_x + \mathbf{e}_z)/\sqrt 2$. For the linearly polarized light, $\mathbf{E}_{LP} = \begin{pmatrix} -\sin\theta/\sqrt 2, & \cos\theta, & \sin\theta/\sqrt 2 \end{pmatrix}$  and for the circularly polarized light, $\mathbf{E}_{CP} = \begin{pmatrix} +1/2, & +1\sqrt 2, & +1/2 \end{pmatrix}$.
The  density matrix of  light, $\sigma _{mn} = E_m E^\ast _n$ for a linearly-?polarized pump $\sigma_{LP}$ has the form
\begin{subequations}
\begin{align}
\sigma_{LP} &=
\begin{pmatrix}
\frac{1}{2}\sin^2\theta  & -\frac{1}{2\sqrt 2} \sin 2\theta & -\frac{1}{2}\sin^2\theta \\ 
-\frac{1}{2\sqrt 2} \sin 2\theta & \cos^2\theta & \frac{1}{2\sqrt 2} \sin 2\theta \\ 
-\frac{1}{2}\sin^2\theta & \frac{1}{2\sqrt 2}\sin 2\theta & \frac{1}{2}\sin^2\theta 
\end{pmatrix}
\end{align}
\end{subequations}
and similarly for $\sigma_{CP}$
\begin{subequations}
\begin{align}
\sigma_{CP} &=
\begin{pmatrix}
1/4 & 1/(2\sqrt 2) & 1/4\\
1/(2\sqrt 2) & 1/2 & 1/(2\sqrt 2)\\
1/4 & 1/(2\sqrt 2) & 1/4
\end{pmatrix}.\label{eq_sigmaCP}
\end{align}
\end{subequations}

Note that $\sigma_{LP}(\theta=45^\circ)$ is very similar to $\sigma_{CP}$:
\begin{subequations}
\begin{align}
\sigma_{LP}(45^\circ) &=
\begin{pmatrix}
1/4 & -1/(2\sqrt 2) & -1/4\\
-1/(2\sqrt 2) & 1/2 & 1/(2\sqrt 2)\\
-1/4 & 1/(2\sqrt 2) & 1/4
\end{pmatrix}.\label{eq_sigma45}
\end{align}
\end{subequations}

The diagonal and nondiagonal elements of the matrix are responsible for the longitudinal and transverse pumping, respectively. The sums $(\sigma_{+0}+\sigma_{0+})$ and $(\sigma_{0-}+\sigma_{-0})$ describe the transverse orientation, while the differences $(\sigma_{+0} - \sigma_{0+})$ and $(\sigma_{0-} - \sigma_{-0})$ describe the transverse light-induced alignment. The diagonal elements are proportional to  intensities of the corresponding components $I_0 =\sigma_{00} =\cos^2\theta$ and $I_+ = I_- = \sigma_{++} = (1/2)\sin^2\theta$. 
  
The longitudinal alignment of the atomic ensemble is created by the longitudinal alignment of the light of approximately  
$I_{+} + I_{-} - 2 I_0 = 1 - 3\cos^2\theta$.
The sign of the alignment depends on the angle $\theta$. The light polarized at the angle $\theta_0$, when $\cos^2\theta _0 = 1/3$, will not create any longitudinal alignment.

Depopulation pumping occurs if excitation probabilities are different for different levels. For the linearly polarized pumping, the probability $P_{Fm_F}$ of the light absorption, from the level $F,m_F$, in the general case, is: 
\begin{equation}
P_{Fm_F}=\sum_{F',m_{F'}} \mathcal{J}_{{FF'}} I_q \frac{2J+1}{2I+1} 
\begin{Bmatrix} J & J' & 1 \\ F' & F & I\end{Bmatrix}^2
\bigl[ C^{F'm_{F'}}_{Fm_F1q} \bigr]^2.
\end{equation}
Here $\mathcal{J}_{{FF'}}$ are the intensities of the transition between the hyperfine level of the ground ($F$) and excited ($F'=F-1,F,F+1$) states, which are the same in the case of an unresolved excitation line; $ m_F$ and $m_{F'}$ are related to magnetic sublevels; $q = m_{F'} - m_F = 0, \pm 1$; $I_q$ are the intensities of the circular components; $C_{Fm_F1q}^{F' m_{F'}}$ is the Clebsch-Gordon coefficient; $\begin{Bmatrix} J & J' & 1 \\ F' & F & I\end{Bmatrix}$ is the Racah 6-$j$ symbol, $L$ ($L'$) and $J$ ($J'$) are the orbital and total angular momenta of the ground (excited) state, respectively; $S=1/2$ is the electron spin and $I = 7/2$ is the nuclear spin. This expression is valid for any light polarization.\looseness=-1

For the linearly polarized pumping, $I_{+1} = I_{-1}$. The longitudinal alignment of level $F$ is calculated by summing excitation probabilities of the Zeeman levels $mF$ with weights of approximately $3m_F^2-F(F+1)$. The alignment of the $F = 3$ level is $A_{3}\sim (-16\mathcal{J}_{32}+21\mathcal{J}_{33}-5\mathcal{J}_{34})\cdot (3\cos^2\theta-1)$. The alignment vanishes when the intensities $\mathcal{J}_{3F'}$ are equal to each other. The alignment at the $F = 4$ level is $A_{4}\sim (-5\mathcal{J}_{43}+21\mathcal{J}_{44}-16\mathcal{J}_{45})\cdot (3\cos^2\theta-1)$. Total alignment also vanishes for equal intensities. 

In the case of unresolved excitation, the optical pumping is performed isotropically and the alignment due to depopulation does not occur in the ground state. However, in the excited states $F'$, the alignment does appear. This is the reason for  the appearance of alignment in the ground state of the vacuum cell. If excitation is provided by the resolved line, the alignment in the ground state by depopulation pumping $F \to F'= F \pm 1$ proves to be opposite in sign  to the alignment by  the $F \to F'= F$ transition. The three contributions to the alignment compensate for each other in the  case of an unresolved line.

In the buffer-gas cells, the spin orientation in the excited state is being lost. The alignment depolarization cross section (as well as orientation) in the excited state $P_{3/2}$ is very high (approximately $1.3 \cdot 10^{-14}$\ cm$^2$ for Cs-Ne collisions \cite{Krause}). Therefore, we may assume that the alignment is destroyed with each collision with the buffer gas. For the cells with buffer gas, the relaxation rate of alignment, for the neon pressure exceeding $1$\ Torr, is higher than the rate of spontaneous decay and the transfer of alignment from the excited state does not occur.

Let us discuss the orientation case in a gas cell. For the circularly polarized pumping, the pumping rates from the Zeeman levels are not the same. For example, the light absorption  probabilities from $m_F = -4, -3, \ldots, +4$ Zeeman levels of $F = 4$ relate to each other as $4:5:6:7:8:9:10:11:12$, so the levels will be emptied unequally. Even if the excited-state orientation is lost in a collision, the ground-state polarization will be provided by the depopulation pumping. In the vacuum cells, the polarization occurs mainly by orientation transfer through the excited states. This difference between the pumping cycles of orientation and alignment gives rise to  their different sensitivity to the presence of buffer gas.

To completely describe the transverse pumping, one has to take into account the effects of the light-induced coherence. The light-intensity modulation excites oscillations of non-diagonal elements of the atomic density matrix, giving rise to resonances of the CPT type. The contribution to the coherence of the ground-state density matrix $\rho_{ik}$ between the Zeeman sublevels $i$ and $k$ is provided by  interference of several $\Lambda$\ schemes involving up to six Zeeman levels of the upper hyperfine states. The resonances of CPT  compensate for each other in the case of an unresolved excitation line, in contrast to the circularly polarized pumping in a gas cell.

The absence of the alignment signal is justified when transitions to the hyperfine levels of the excited state are not resolved. For the sodium,  potassium, or rubidium atoms pumped by the $D_2$ line, this condition is definitely satisfied, since the Doppler width substantially exceeds the hyperfine splitting of the excited state. For cesium, the Doppler half-width is close to the hyperfine splitting of the excited state. When tuning the laser frequency to the wing of the absorption line, e.g., to the short-wavelength region of the $F = 4 \to F' = 5$ transition, one of the transitions can be predominantly excited. In this case, the alignment in the ground state can also be created in the gas cell. A more detailed study of the excitation spectrum of the spin alignment is beyond the scope of the present paper.  

\section{Conclusion}

Our experimental results show that the signal of the spin-alignment precession driven by the linearly polarized light at the Larmor frequency, in vacuum cells, is practically equal in magnitude to that of the spin-orientation precession and therefore can be equally used efficiently as a tool of physical research and magnetometric applications. As known, the spin-alignment precession (unlike precession of spin orientation) is not accompanied by the precession of magnetization that makes it insensitive to the resonant magnetic field and does not allow one to use the oscillating magnetic field for its excitation or detection. This fact can be considered, depending on the type of problem to be solved, either as a drawback or as a merit of the method. 

In spite of this tolerance of the spin alignment to magnetic perturbations, the optically driven spin alignment proved to be  sensitive (essentially more sensitive than the spin orientation) to collisions with rare-gas atoms, which are commonly used to slow down atomic diffusion with no loss of spin orientation. We believe that this effect, which may seem paradoxical, can be explained by the fact that the linearly polarized light does not produce the depopulation-type pumping due to equality of the excitation probabilities, for a nonresolved hyperfine structure of the excited state. It is important that the creation of alignment in a cell with gas by means of nonequilibrium population of levels from the excited state is excluded because of the complete destruction of alignment in the excited state. In the vacuum cell, the alignment arises due to population of the ground-state sublevels through  transfer of the alignment from the excited state. 

We believe that the results of this work will advance the use of the effect of spin-alignment precession in fundamental and applied research.

\acknowledgements
We greatly appreciate financial support for the experimental work from the Russian Science Foundation (Grant No. 21-72-10021). A.S.P. acknowledges support for the theoretical work from the RFBR Grant No. 19-52-12054. The samples were provided via support from the St. Petersburg State University Research Grant No. 94030557.

\end{document}